\documentclass[12pt,a4paper]{article}

\usepackage{amsmath,amssymb,epsfig}
\begin{document}
\numberwithin{equation}{section}
\setlength{\unitlength}{.8mm}

\begin{titlepage} 
%\begin{flushright}
%version1\\
%\today
%\end{flushright}
\vspace*{0.5cm}
\begin{center}
{\Large\bf An exact solution of the Currie-Hill equations in $1 + 1$ 
dimensional Minkowski space}
\end{center}
\vspace{2.5cm}
\begin{center}
{\large J\'anos Balog }
\end{center}
\bigskip
\begin{center}
{\it MTA Lend\"ulet Holographic QFT Group, Wigner Research Centre,\\
H-1525 Budapest 114, P.O.B. 49, Hungary}
\end{center}
\vspace{3.2cm}
\begin{abstract}
%\noindent
We present an exact two-particle solution of the Currie-Hill equations 
of Predictive Relativistic Mechanics in $1 + 1$ dimensional Minkowski space.
The instantaneous accelerations are given in terms of elementary functions
depending on the relative particle position and velocities. The general 
solution of the equations of motion is given and by studying the global
phase space of this system it is shown that this is a subspace of the full 
kinematic phase space.
\end{abstract}

\end{titlepage}

\section{Introduction}

Relativistic particle mechanics, with instantaneous action-at-a-distance 
interaction naively contradicts relativistic causality and is thus 
counter-intuitive. Despite of this apparent difficulty, a consistent theory 
exists, mainly in the classical domain, but also quantum-mechanically. 
Nevertheless, relativistic particle physics remains almost completely 
synonymous with Relativistic Quantum Field Theory. Abandoning the particle 
alternative is partially due to the famous no-go theorem of Currie, 
Jordan and Sudarshan \cite{1}. This theorem states
that requiring particle positions to satisfy canonical commutation 
relations excludes the presence of any non-trivial interactions.
If we give up this requirement, we can formulate relativistic 
point mechanics. There are three, essentially equivalent approaches. 
The first one is called Predictive Relativistic Mechanics (PRM) \cite{2} 
and is formulated by writing the equations of motion in Newtonian form
\begin{equation}
\ddot{x}^i_a=A^i_a(\{x\},\{\dot{x}\}),
\label{Newton}
\end{equation}
where $i = 1, 2, 3$ are space indices, $a = 1, 2,\dots,N$ are particle 
indices and the accelerations $A^i_a$
occurring in the Newton-equations (\ref{Newton}) depend
on the instantaneous positions $x^i_a$ and velocities $\dot{x}^i_a$
of the particles. Relativistic invariance
implies that the accelerations have to satisfy a set of quadratic, 
partial differential equations, the Currie-Hill (CH) equations \cite{3}:
\begin{equation}
\begin{split}
\sum_b\Big\{\frac{\partial A^i_a}{\partial v^k_b}
&+\frac{1}{c^2}(x^k_a-x^k_b)v^j_b\,\frac{\partial A^i_a}{\partial x^j_b}
+\frac{1}{c^2}(x^k_a-x^k_b)A^j_b\,\frac{\partial A^i_a}{\partial v^j_b}
-\frac{1}{c^2}v^k_bv^j_b\,\frac{\partial A^i_a}{\partial v^j_b}\Big\}\\
&+\frac{2}{c^2}v^k_aA^i_a+\frac{1}{c^2}A^k_av^i_a=0.
\end{split}
\label{CH31}
\end{equation}
Here $c$ is the speed of light and we introduced the notation 
$v^i_a=\dot{x}^i_a$ and Einstein summation
convention is used for the (upper) space indices $i$, $j$, $k$ but not for
particle (lower) indices $a$, $b$. The Currie-Hill equations 
ensure that if we transform the Newton equations to a
Lorentz-boosted new coordinate system, the particle trajectories satisfy
Newton equations which are instantaneous action-at-a-distance equations
in the boosted coordinate system and moreover the equations in the new system
are of the same form as (\ref{Newton}).

One of the difficulties of the relativistic particle dynamics is that
unfortunately no explicit solution of the Currie-Hill equations is known, 
neither in $3 + 1$ space-time dimensions nor in $1 + 1$
dimensions. Although the most general 2-particle solution has been found 
in $1 + 1$ dimensions \cite{4}, but it was given in a very implicit form. 
There exist approximate solutions in the $1/c^2$ expansion but the absence 
of explicit exact solutions make the study of further questions like the
global structure of the phase space, symplectic structure, etc. difficult.
In this paper we are presenting a completely explicit solution of
the Currie-Hill equations in $1 + 1$ dimensional Minkowski space, 
written in terms of elementary functions. This explicit solution provides 
an example in which further questions of the relativistic 
action-at-a distance approach (conserved quantities, canonical structure, 
etc.) can be studied transparently.

Having found a solution of the Currie-Hill
equations the next natural question is about the existence 
(and uniqueness) of the 10 integrals corresponding to the Poincar\'e group. 
If these exist then one can ask further if a symplectic structure on the 
phase space (the space of all solutions) can
be constructed such that these 10 integrals generate the Poincar\'e group.
An alternative approach to relativistic mechanics \cite{5} 
can be called canonical. Here a phase space equipped with a symplectic 
structure is assumed from the beginning, together with the set of 
10 generators of the Poincar\'e group. In this approach
consistent relativistic dynamics can be constructed if we 
can find the particle positions  $x^i_a$, as functions on the phase 
space and satisfying the Poisson-bracket relations
\begin{equation}
\{ P^i,x^j_a\}=\delta^{ij},\qquad
\{ J^i,x^j_a\}=\epsilon^{ijk}x^k_a,\qquad
\{ K^i,x^j_a\}=\frac{1}{c^2}x^i_a\dot{x}^j_a
\label{WLC}
\end{equation}
called the world line conditions. Here $P^i$, $J^i$, $K^i$, 
respectively are the momentum, angular momentum, and Lorentz boost generators,
respectively, of the Poincar\'e group. If we are able to find such 
particle coordinates, we can calculate the Poisson brackets
\begin{equation}
\{ x^i_a,x^j_b\},
\label{CR}
\end{equation}
which must not vanish, otherwise, due to the no-go theorem, 
there is no interaction. The advantage of the canonical approach is that 
only the coordinates have to be constructed, the 10 integrals of the 
Poincar\'e group are there by construction from the beginning. 
Provided the set $\{x^i_a\}$, $\{\dot{x}^i_a\}$ 
are good coordinates on the phase space (at least locally), the accelerations 
in the Newton-equations (\ref{Newton}) can be calculated 
and must satisfy the Currie-Hill equations. There is also a third, 
essentially equivalent approach \cite{6} which is  explicitly 
covariant. This is not discussed here.

A physical example for relativistic particle interactions
is provided by the classical electrodynamics 
of point charges \cite{7} either in the Feynman-Wheeler formulation 
or as in Rohrlich's theory. The equations
of motion are only known in the post-Coulombian expansion 
(where the expansion parameter is $1/c^2$). The problem of classical 
electrodynamics of point charges may be academic, but it is a 
somewhat simpler analog of the physically relevant problem of 
motion of compact binaries in general relativity (modeling the bound
states of two black holes or two neutron stars). In the latter 
case the equations of motion are known up to the $3^{\rm rd}$ post-Newtonian 
order (up to the terms proportional  to $c^{-6}$) \cite{8} and they satisfy 
the Currie-Hill equations (in the post-Newtonian perturbative
sense). It is not clear if the expansion can be extended further 
(the system starts radiating gravitational waves at the $2.5^{\rm th}$ 
post-Newtonian order).

Because of the lack of explicit exact solutions it is 
important to study $1 + 1$ dimensional examples, the most famous of which are
the exactly solvable 
Ruijsenaars-Schneider (RS) models \cite{9}, the relativistic generalizations 
of the Calogero-Moser systems. The RS approach is canonical, and
the RS systems are not only relativistic, but also integrable 
for any $N$. The original motivation of constructing the RS models 
was their relativistic invariance but later
the RS literature was almost entirely concerned with their 
integrable aspects (there are many applications of the RS 
models in various areas of physics). Here trajectory variables 
satisfying the $1+1$ dimensional version of
the word line conditions (\ref{WLC}) have been constructed 
but it is not clear if they are good coordinates on the entire phase
space and their explicit form in terms of the canonical variables
and their commutation relations (\ref{CR}) are not known explicitly.
There are also further open questions even in the case of RS models
(the question of physical non-relativistic limit, for instance) and
for this reason it is important to study further examples where all
physical questions can be studied more easily. The example we
are presenting here is simple enough to do further calculations and
to study global questions effortlessly.

We will present our $1 + 1$ dimensional solution in the next section.
We will construct the conserved quantities associated to the Poincar\'e
generators in section~3.
Some conclusion and a list of further questions which can be studied
using this example is discussed in the Conclusion section.

\section{A solution of the Currie-Hill equations in $1 + 1$ dimensions}

In this section we will solve the CH equations in $1 + 1$ dimensions
for $N=2$ particles. Rescaling some of the variables we introduce
\begin{equation}
y=x_1-x_2,\qquad u_a=\frac{1}{c}v_a,\qquad
A_a=c^2\omega_a(y,u_1,u_2),\quad (a=1,\, 2).
\end{equation}
In terms of the new variables the CH equations simplify:
\begin{equation}
\begin{aligned}
(1-u_1^2)\,\frac{\partial\omega_1}{\partial u_1}
+(1-u_2^2+y\omega_2)\,\frac{\partial\omega_1}{\partial u_2}
-yu_2\,\frac{\partial\omega_1}{\partial y}+3u_1\omega_1&=0,\\
(1-u_1^2-y\omega_1)\,\frac{\partial\omega_2}{\partial u_1}
+(1-u_2^2)\,\frac{\partial\omega_2}{\partial u_2}
-yu_1\,\frac{\partial\omega_1}{\partial y}+3u_2\omega_2&=0.
\end{aligned}
\label{CH11}
\end{equation}

Not all solutions of the CH equations are physically acceptable
in relativistic mechanics. One of the missing ingredients is the 
relativistic generalization of Newton's third law (action--reaction).
In a nonrelativistic two-particle problem we would
require (in addition to the Galilean version of (\ref{CH11}))
\begin{equation}
m_1A_1=-m_2A_2,
\label{NIII}
\end{equation}
where $m_1$ and $m_2$ are the masses of the particles. As is well known
this is equivalent to the statement that the centre of mass of the two-particle
system is moving uniformly. In the absence of a proper generalization of 
the notion of centre of mass for relativistic particles and that of the
third law we restrict our attention here to the case of two identical 
particles. In this case, by symmetry considerations we can assume that
$(x_1+x_2)/2$ moves uniformly and we add to (\ref{CH11}) the requirement
\begin{equation}
\omega_1=-\omega_2.
\end{equation}

Even after this simplification the CH equations (\ref{CH11}) are complicated
nonlinear partial differential equations. Hill \cite{4} found the general 
solution of the equations in an implicit form. 
Although locally it provides the general
solution in terms of two arbitrary functions of two variables, the implicit 
nature of the solution makes the investigation of global questions difficult.
On the other hand, we will see that the particular solution presented in 
this paper is more suitable for global considerations. 

We will look for solutions where the accelerations
depend only on the combination
\begin{equation}
\xi=\frac{1-u_1u_2}{y}
\end{equation}
of the kinematic variables. It turns out that using the Ansatz
\begin{equation}
\omega_1=-\omega_2=f(\xi)
\end{equation}
both equations in (\ref{CH11}) reduce to the nonlinear ordinary differential
equation
\begin{equation}
(f-\xi)f^\prime+3f=0.
\label{fxi}
\end{equation}

To solve (\ref{fxi}) we first of all write
\begin{equation}
f=\frac{4}{\ell}\,h^{3/2},
\label{hpara}
\end{equation}
where $h$ is a new function characterizing the accelerations and $\ell$ is
our unit of length (it could be scaled out from the problem). In terms of
$h$, (\ref{fxi}) takes the form
\begin{equation}
\frac{4}{\ell}\,h^{3/2}h^\prime-\xi h^\prime+2h=0.
\label{hxi}
\end{equation}

We now present the solution of (\ref{hxi}) assuming that
\begin{equation}
0<h<1.
\end{equation}
The solution is given implicitly by
\begin{equation}
h(h-1)^2=\frac{\ell^2}{4}\,\xi^2.
\label{sol}
\end{equation}
This solution can be made explicit by expressing $h$ in terms of $\xi$
using Cardano's formula. Moreover, the result is an elementary, algebraic
function built from square and cubic roots\footnote{
$h=\frac{2}{3}+\sqrt[3]{\frac{Z}{2}-\frac{1}{27}+\frac{i}{2}\sqrt{Z\left(
\frac{4}{27}-Z\right)}}
+\sqrt[3]{\frac{Z}{2}-\frac{1}{27}-\frac{i}{2}\sqrt{Z\left(
\frac{4}{27}-Z\right)}}$
}. 
It is, however, easier to
study its properties using (\ref{sol}) in its original form. Both the
variable $\xi$ and the derivative $h^\prime$ can be expressed using
(\ref{sol}) and its derivative:
\begin{equation}
\xi=\frac{2}{\ell}\,\sqrt{h}(1-h),\qquad\quad 
h^\prime=\frac{\ell\sqrt{h}}{1-3h}.
\end{equation}
Putting these expressions to (\ref{hxi}) we see that it is indeed satisfied.

To summarize, we have to find a solution of
\begin{equation}
h(1-h)^2=Z,\qquad\quad Z=\left(\frac{\ell(1-u_1u_2)}{2y}\right)^2
\end{equation}
in the range $0<h<1$ and this parametrizes the acceleration (\ref{hpara}).
The function $Z=h(1-h)^2$ has a single maximum in this range (between $0$
and $1$) at $h=1/3$. The maximum value is $4/27$.
Thus there is no solution unless
\begin{equation}
0<Z<\frac{4}{27},\qquad\qquad y>\frac{3\sqrt{3}\ell(1-u_1u_2)}{4}.
\label{ineq0}
\end{equation}
This is a new feature of relativistic mechanics. In contrast to Newtonian
mechanics, here the initial conditions are not arbitrary. We have to require
that the initial conditions satisfy (\ref{ineq0}) in addition to the obvious
\begin{equation}
|u_1|<1,\qquad\qquad |u_2|<1.
\end{equation}
If $Z$ is in the allowed range, there are two solutions for $h$. We will
call the 
\begin{equation}
0<h<\frac{1}{3}
\end{equation}
solution the \lq\lq good'' branch. The mapping between $Z$ and $h$ in the
\lq\lq good'' branch is one-to-one if (\ref{ineq0}) is satisfied.

More detailed considerations reveal that (\ref{ineq0}) is only a necessary
condition and the phase space should be restricted further. The reason is that
it is possible that even if we start from a phase space point satisfying 
(\ref{ineq0}), during the later (or earlier) time evolution of the system 
we leave this part of the phase space and the accelerations are no longer 
well defined. We should find a smaller subspace of the full kinematic
phase space with the property that starting from here the entire future and 
past time evolution of the system remains within this subspace. To investigate
the global structure of the phase space thus requires the knowledge of the
solution of equations of motion, which makes this analysis difficult in 
general. For the case at hand, the solution of the equations of motion is
available and we found that the necessary and sufficient condition for
the existence of a uniquely determined and globally well defined (in the 
above sense) pair of particle trajectories which go through the given phase 
space point is
\begin{equation}
h_o>0,\qquad\qquad y>\frac{\ell(1-u_1u_2)}{2\sqrt{h_o}(1-h_o)}.
\label{ineq}
\end{equation}
Here $h_o$ depends only on the velocities and is given by
\begin{equation}
p(1-h_o)=1+g+\sqrt{g^2+\frac{1+2g}{9}},
\end{equation}
where
\begin{equation}
g=\frac{u_1+u_2}{2-u_1-u_2},\qquad\qquad
p=\frac{2+u_1+u_2}{1-u_1u_2}.
\end{equation}
It can be shown that
\begin{equation}
h_o\leq\frac{1}{3}.
\end{equation}
The second requirement in (\ref{ineq}) comes from the \lq\lq good'' branch
condition $0<h<h_o$.

To better understand the meaning of (\ref{ineq}) and to simplify the formulas
we now go to the centre of mass system
\begin{equation}
u_1=-u_2=u.
\label{comu}
\end{equation}
Here
\begin{equation}
g=0,\qquad\qquad p=\frac{2}{1+u^2}
\end{equation}
and the physical subspace is given by
\begin{equation}
u^2<\frac{1}{2},\qquad\qquad y>\frac{3\sqrt{3}\ell}{4\sqrt{1-2u^2}}.
\end{equation}

The general solution of the equations of motion for our system
can be given in algebraic form as function of the time $t$ and integration 
constants. In the centre of mass system\footnote{The problem of defining the 
centre of mass of the system is discussed in section 3. Here we just use the 
coordinate system defined by (\ref{comu}) and (\ref{comx}).} 
where
\begin{equation}
x_1=-x_2=\frac{y}{2}
\label{comx}
\end{equation}
the solution becomes very simple and is of the form
\begin{equation}
y=2b\sqrt{t^2+B},\qquad\qquad u=\frac{bt}{\sqrt{t^2+B}}.
\label{tra}
\end{equation}
There is a single integration constant\footnote{The other, less relevant 
integration constant is fixed by identifying the origin of the time coordinate
with the turning point of the scattering process.} 
$A>1$ in this case and the
constants $b$ and $B$ are
\begin{equation}
b=\sqrt{\frac{A-1}{A+1}},\qquad\qquad
B=\frac{\ell^2A^3}{2(A+1)(A-1)^2}.
\end{equation}
We see that the solution describes a time-reversal symmetric scattering 
process in which the two particles repel each other. They come in from 
infinity, gradually approach 
each other and after the turning point, where the particles stop and reach the
minimal relative distance, are receding from each other.

We can also calculate
\begin{equation}
h=\frac{A-1}{2A}\,\frac{B}{t^2+B}
\end{equation}
and we see that
\begin{equation}
0<h<h_{\rm max},
\end{equation}
where
\begin{equation}
h_{\rm max}=\frac{A-1}{2A}.
\end{equation}

The trajectories given by (\ref{tra}) exist for any $A>1$ but the
\lq\lq good'' branch condition is only satisfied for
\begin{equation}
1<A<3,\qquad\qquad h_{\rm max}<\frac{1}{3}.
\end{equation}

Finally we note that in the asymptotic past
\begin{equation}
t\to -\infty\qquad\qquad x_1^{(-\infty)}=-x_2^{(-\infty)}\approx -bt
\end{equation}
and similarly
\begin{equation}
t\to +\infty\qquad\qquad x_1^{(+\infty)}=-x_2^{(+\infty)}\approx bt,
\end{equation}
which means that the process is not very interesting from the point of view
of scattering theory. Although the asymptotic
velocities are swapped between the particles, this is a \lq\lq billiard ball''
type scattering, where the time delay vanishes.

\section{Construction of the Poincar\'e generators}

In this section we construct the conserved quantities associated to the
three generators of the $1+1$ dimensional Poincar\'e group. We first 
describe the general strategy of the construction
and then carry out the calculation explicitly for our special case.

The three Poincar\'e generators on the phase space are represented by
the differential operators $\hat{\cal H}$, $\hat{\cal P}$ and $\hat{\cal K}$.
These are respectively the generators of the time translation, space
translation and Lorentz boost. The corresponding functions on the phase
space are respectively the Hamiltonian ${\cal H}$, the momentum ${\cal P}$
and the (rescaled) centre of mass ${\cal K}$. The generators satisfy the
commutation relations of the Poincar\'e Lie algebra:
\begin{equation}
[\hat {\cal H},\hat{\cal P}]=0,\qquad\quad
[\hat {\cal H},\hat{\cal K}]=\hat{\cal P},\qquad\quad
[\hat {\cal P},\hat{\cal K}]=\frac{1}{c^2}\hat{\cal H}.
\label{PLie}
\end{equation}
In a canonical mechanical system the Poincar\'e transformations on phase
space functions ${\cal F}$ would be generated via the Poisson bracket 
relations
\begin{equation}
\hat A{\cal F}=\{A,{\cal F}\},\qquad\qquad[\hat A,\hat B]=\widehat{\{A,B\}}.
\label{PB}
\end{equation}
The same Poisson bracket relations are the representation
of the Poincar\'e Lie algebra and at the same time the transformation
rules of the quantities ${\cal H}$, ${\cal P}$, ${\cal K}$ under
infinitesimal Poincar\'e transformations. Here we have no canonical structure
but can read off the latter rules by combining (\ref{PLie}) and (\ref{PB}):
\begin{equation}
\begin{split}
\hat{\cal P}{\cal H}&=\hat{\cal H}{\cal H}=0,\qquad\quad
\hat{\cal K}{\cal H}=-{\cal P},\\
\hat{\cal P}{\cal P}&=\hat{\cal H}{\cal P}=0,\qquad\quad
\hat{\cal K}{\cal P}=-\frac{1}{c^2}{\cal H},\\
\hat{\cal K}{\cal K}&=0,\qquad\quad \hat{\cal H}{\cal K}={\cal P},\qquad\quad
\hat{\cal P}{\cal K}=\frac{1}{c^2}{\cal H}.
\end{split}
\label{Trules}
\end{equation}
From the structure of the above set of transformation rules and the commutation
relations (\ref{PLie}) we can see that a possible strategy of construction
is to find a suitable ${\cal K}$ satisfying
\begin{equation}
\hat{\cal K}{\cal K}=
\hat{\cal P}\hat{\cal H}{\cal K}=
\hat{\cal H}^2{\cal K}=
\hat{\cal P}^2{\cal K}=0.
\label{Keqs}
\end{equation}
All the relations (\ref{Trules}) are satisfied if we complete the set of
conserved quantities by
\begin{equation}
{\cal P}=\hat{\cal H}{\cal K},\qquad\quad
{\cal H}=c^2\hat{\cal P}{\cal K}.
\end{equation}

In a $1+1$ dimensional predictive relativistic system the local
coordinates on the phase space are the particle positions $x_a$ and
velocities $v_a$ and the Poincar\'e generators are represented by
\begin{equation}
\begin{split}
\hat{\cal P}&=-\sum_a\frac{\partial}{\partial x_a},\\
\hat{\cal H}&=\sum_a\left\{v_a\frac{\partial}{\partial x_a}
+A_a\frac{\partial}{\partial v_a}\right\},\\
\hat{\cal K}&=\sum_a\left\{-\frac{x_av_a}{c^2}\frac{\partial}{\partial x_a}
+\left(1-\frac{v_a^2}{c^2}-\frac{x_aA_a}{c^2}\right)
\frac{\partial}{\partial v_a}\right\}.
\end{split}
\label{diff}
\end{equation}
The above representation implements the $1+1$ dimensional version of the
world line conditions (\ref{WLC}) and in fact the Currie-Hill equations
(\ref{CH11}) are nothing but the requirement that the differential operators
(\ref{diff}) satisfy the algebra (\ref{PLie}).

In the case of a single free particle there is no need for the particle
label and the differential operators are
\begin{equation}
\hat{\cal P}=-\frac{\partial}{\partial x},\qquad\quad
\hat{\cal H}=v\frac{\partial}{\partial x},\qquad\quad
\hat{\cal K}=-\frac{xv}{c^2}\frac{\partial}{\partial x}
+\left(1-\frac{v^2}{c^2}\right)\frac{\partial}{\partial v}.
\end{equation}
In this case it is easy to find the general solution of the differential
equations (\ref{Keqs}) and in this way we reproduce the familiar formulas
\begin{equation}
{\cal K}=-\frac{mx}{\sqrt{1-\frac{v^2}{c^2}}}+\beta_o,\qquad
{\cal H}=\frac{mc^2}{\sqrt{1-\frac{v^2}{c^2}}},\qquad
{\cal P}=-\frac{mv}{\sqrt{1-\frac{v^2}{c^2}}},
\end{equation}
where $m$ and $\beta_o$ are constants. The physical meaning of $m$ is obvious
while we can set $\beta_o=0$ by requiring parity invariance. The physical
meaning of the  conserved quantities\footnote{Of course the 
centre of mass is not conserved since its time derivative is given by the total
momentum but we will continue to call the set $\{{\cal H},{\cal P},{\cal K}\}$
\lq\lq conserved'' quantities.} is here, and in general,
\begin{equation}
{\cal H}=E,\qquad\quad
{\cal P}=-P,\qquad\quad
{\cal K}=-\frac{EY}{c^2},\qquad\quad
\end{equation}
where $E$ is the total energy, $P$ the total momentum and $Y$ the centre of 
mass. Note the presence of some minus signs which are due to our conventions.

Next we discuss the case of two symmetric particles where
\begin{equation}
A_1=-A_2=f.
\end{equation}
From now on in the rest of this section to simplify the formulas we will 
set $c=1$ and use the variables
\begin{equation}
x_1+x_2=X,\qquad\quad
x_1-x_2=y,\qquad\quad
v_1+v_2=w,\qquad\quad
v_1-v_2=v.
\end{equation}
The generators are
\begin{equation}
\hat {\cal P}=-2\frac{\partial}{\partial X},\qquad\quad
\hat {\cal H}=w\frac{\partial}{\partial X}+v\frac{\partial}{\partial y}
+f\left(\frac{\partial}{\partial v_1}-\frac{\partial}{\partial v_2}\right)
\end{equation}
and
\begin{equation}
\begin{split}
\hat{\cal K}=&-\frac{Xw+yv}{2}\frac{\partial}{\partial X}
-\frac{Xv}{2}\frac{\partial}{\partial y}-\frac{Xf}{2}\left(
\frac{\partial}{\partial v_1}-\frac{\partial}{\partial v_2}\right)-
\frac{yw}{2}\frac{\partial}{\partial y}\\
&+(1-v_1^2)\frac{\partial}{\partial v_1}
+(1-v_2^2)\frac{\partial}{\partial v_2}
-\frac{yf}{2}\left(
\frac{\partial}{\partial v_1}+\frac{\partial}{\partial v_2}\right).
\end{split}
\end{equation}

We now present the general solution of the differential equations (\ref{Keqs})
in the special case discussed in this paper. In this special case $f=f(\xi)$
and is given by (\ref{hpara}) and (\ref{sol}). In this section we choose our 
unit of length so that we can set $\ell=2$. We now define the variables
\begin{equation}
\varepsilon=y(2\xi+f),\qquad\quad \Gamma=w^2+2(\varepsilon -2),\qquad\quad
T=\frac{yv}{\Gamma},\qquad\quad q=\frac{\Gamma}{\varepsilon^2}.
\end{equation}
These satisfy
\begin{equation}
\hat{\cal H}\varepsilon=0,\qquad \hat{\cal H} T=1,\qquad\quad
\hat {\cal H}q=\hat{\cal P}q=\hat{\cal K}q=0.
\end{equation}
$\varepsilon$, $w$ and $\Gamma$ are time-independent and translation
invariant, while $q$ is Poincar\'e invariant. Using these new variables,
the general solution is of the form
\begin{equation}
{\cal K}=AX+DT+B,\qquad\quad{\cal H}=-2A,\qquad\quad{\cal P}=Aw+D,
\end{equation}
where
\begin{equation}
B=B(q),\qquad\quad A=\frac{1}{\sqrt{\varepsilon}}\,g(\varepsilon,q),\qquad\quad
D=-2w\sqrt{\varepsilon}\,\frac{\partial g}{\partial\varepsilon}
\end{equation}
and $g(\varepsilon,q)$ has to satisfy the second order (ordinary)
differential equation
\begin{equation}
qg=4(q\varepsilon^2-2\varepsilon+4)\frac{\partial^2 g}{\partial\varepsilon^2}
+4(q\varepsilon-1)\frac{\partial g}{\partial\varepsilon}.
\end{equation}
The latter has general solution
\begin{equation}
g=g_1(q){\cal R}_++g_2(q){\cal R}_-,
\end{equation}
where
\begin{equation}
{\cal R}_\pm=\sqrt{\frac{1}{q}-\varepsilon\pm\frac{\sqrt{1-4q}}{q}}.
\end{equation}

The physical meaning of the conserved quantities can be better understood 
if we introduce the asymptotic rapidities of the particles. Since as we 
have seen in the preceding section our system describes the scattering 
of the two particles, in the asymptotic past we have
\begin{equation}
t\to-\infty\qquad\qquad v_1\to\tanh\beta_1,\qquad\quad v_2\to\tanh\beta_2.
\end{equation}
We introduce the combinations
\begin{equation}
2\beta=\beta_1+\beta_2,\qquad\qquad 2\theta=\beta_2-\beta_1.
\end{equation}
(Note that the interaction is repulsive and the phase space can be reduced
to $y=x_1-x_2>0$ and $2\theta=\beta_2-\beta_1>0$.) Physical meaning of the
conserved quantities can be assessed using the formulas
\begin{equation}
\varepsilon=\frac{4\cosh2\theta}{\cosh2\theta+\cosh2\beta},\qquad\quad
w=\frac{2\sinh2\beta}{\cosh2\theta+\cosh2\beta},\qquad\quad
q=\frac{1}{4}\tanh^2 2\theta.
\end{equation}
The physical meaning of the solution for energy and momentum is given by
\begin{equation}
{\cal H}=-2g_1\frac{\cosh\beta}{\sinh\theta}-2g_2\frac{\sinh|\beta|}
{\cosh\theta},\qquad\quad
{\cal P}=2g_1\frac{\sinh\beta}{\sinh\theta}+2g_2\frac{ 
{\rm sign}(\beta)\cosh\beta}
{\cosh\theta}
\end{equation}
and shows that the natural choice is
\begin{equation}
g_1(q)=-\frac{m\sqrt{q}}{\sqrt{1-4q}}=-m\sinh\theta\cosh\theta,\qquad
\qquad g_2(q)=0
\end{equation}
leading to the usual formulas
\begin{equation}
{\cal H}=E=2m\cosh\theta\cosh\beta,\qquad\qquad
{\cal P}=-P=-2m\cosh\theta\sinh\beta
\end{equation}
and
\begin{equation}
{\cal K}=B(q)-mX\cosh\theta\cosh\beta+my\sinh\theta\sinh\beta.
\end{equation}
Further it is natural to require that ${\cal K}=0$ in the centre of mass 
system, where $X=\beta=0$. This means that we have to choose $B(q)=0$. This
must hold in all coordinate systems since this requirement is Poincar\'e 
invariant. Expressing the conserved quantities in terms of the original 
variables, we finally have
\begin{equation}
{\cal H}=2\mu R,\qquad\quad{\cal P}=-\frac{\mu w}{R}[1+\sqrt{1-4q}],
\qquad\quad {\cal K}=-\mu\left[RX+\frac{yvw}{R\varepsilon}\right],
\end{equation}
where
\begin{equation}
\mu=\frac{m}{\sqrt{\varepsilon(1-4q)}},\qquad\qquad
R=\sqrt{1-q\varepsilon+\sqrt{1-4q}}.
\end{equation}

We note that the centre of mass is given by
\begin{equation}
Y=-\frac{\cal K}{\cal H}=\frac{X}{2}+\frac{yvw}{2R^2\varepsilon},
\label{COM}
\end{equation}
which is different from the naive arithmetic mean of the two coordinates.
The latter does not define a proper trajectory (not even for two free 
particles). It is known that the problem of defining the centre of mass 
of relativistic systems is more complicated than the corresponding Newtonian 
case. For a discussion of this problem see ref.~\cite{ACL} and references 
therein. The choice (\ref{COM}) is the $1+1$ dimensional analog of the 
Fokker-Pryce centre of inertia and satisfies
\begin{equation}
\hat{\cal H}Y=V=\frac{P}{E}={\rm const},\qquad\qquad
\hat{\cal P}Y=-1,\qquad\qquad \hat{\cal K}Y=-YV, 
\end{equation}
i.e. the world line conditions for an effective free particle.

\section{Conclusion}

We have constructed a $1 + 1$ dimensional two-particle relativistic scattering
system where the equations of motion can be written in instantaneous
action-at-a-distance form and expressed the accelerations as function
of the relative distance and particle velocities in terms of elementary 
functions. We have seen that an interesting new feature with respect to
nonrelativistic Newtonian scattering is that initial positions and velocities
can not be chosen arbitrarily, the allowed set of initial conditions is a
subspace of the full kinematic phase space only.

The reason for studying toy models like the one here is that we can hope 
to be able to learn something about the unusual features of relativistic point
mechanics, which remain valid for more realistic models as well.
The following is a list of natural questions that can be studied in any
relativistic particle system based on the PRM approach and in particular, 
can be answered for our simple example.

\begin{itemize}

\item
Construct the 10 (3 in the case of $1 + 1$ dimensions) 
conserved quantities of the Poincar\'e algebra. 

\item
Equip the phase space (defined as the solution space) with symplectic
structure such that the above 10 (3) conserved quantities generate the 
Poincar\'e transformations on the phase space and in particular the word 
line conditions (\ref{WLC}) are satisfied.

\item
Calculate the Poisson brackets (\ref{CR}) and see how the no interaction
theorem is circumvented.

\item
See if a Lagrangian (or action) approach is available for the system.

\item
(For scattering problems) calculate the time delay (classical analog of
scattering phase shifts) as function of asymptotic data (asymptotic
momenta of particles).

\end{itemize}
We have answered the first and last questions in the above list for our simple
example and hope to be able to return to the remaining questions in a separate 
publication.

\vspace{0.5cm}
{\tt Acknowledgements}

\noindent 
This investigation was supported by the Hungarian National Science Fund 
OTKA (under K83267).

%\newpage

\end{document}